\documentclass[letterpaper, 12pt]{article}[2000/05/19]
\usepackage[english]{babel}
\usepackage{amsfonts,amsmath,amssymb,amsthm,latexsym,amscd,mathrsfs}
\usepackage{ifthen,cite}
\usepackage[bookmarksnumbered=true]{hyperref}

\hypersetup{pdfpagetransition={Split}}

\newcommand{\evenhead}{Author \ name}
\newcommand{\oddhead}{Article \ name}
\newcommand{\theArticleName}{Article \ name}

\newcommand{\FirstPageHeading}[1]{\thispagestyle{empty}%
\noindent\raisebox{0pt}[0pt][0pt]{\makebox[\textwidth]{\protect\footnotesize \sf }}\par}

\newcommand{\ArticleName}[1]{\renewcommand{\theArticleName}{#1}\vspace{-2mm}\par\noindent {\LARGE\bf  #1\par}}
\newcommand{\Author}[1]{\vspace{5mm}\par\noindent {\Large  #1\par} \par\vspace{2mm}\par}
\newcommand{\Address}[1]{\vspace{2mm}\par\noindent {\it #1} \par}
\newcommand{\Email}[1]{\ifthenelse{\equal{#1}{}}{}{\par\noindent {\rm E-mail: }{\it  #1} \par}}
\newcommand{\URLaddress}[1]{\ifthenelse{\equal{#1}{}}{}{\par\noindent {\rm URL: }{\tt  #1} \par}}
\newcommand{\EmailD}[1]{\ifthenelse{\equal{#1}{}}{}{\par\noindent {$\phantom{\dag}$~\rm E-mail: }{\it  #1} \par}}
\newcommand{\URLaddressD}[1]{\ifthenelse{\equal{#1}{}}{}{\par\noindent {$\phantom{\dag}$~\rm URL: }{\tt  #1} \par}}

\newcommand{\Abstract}[1]{\vspace{6mm}\par\noindent\hspace*{10mm}
\parbox{140mm}{\small {\bf Abstract.} #1}\par}
\newcommand{\Keywords}[1]{\vspace{3mm}\par\noindent\hspace*{10mm}
\parbox{140mm}{\small {\bf Key words:} \rm #1}\par}
\newcommand{\Classification}[1]{\vspace{3mm}\par\noindent\hspace*{10mm}
\parbox{140mm}{\small {\it 2000 Mathematics Subject Classification:} \rm #1}\vspace{3mm}\par}
\newcommand{\ShortArticleName}[1]{\renewcommand{\oddhead}{#1}}
\newcommand{\AuthorNameForHeading}[1]{\renewcommand{\evenhead}{#1}}

\long\def\@makecaption#1#2{
  \sbox\@tempboxa{\small \textbf{#1.}\ \ #2}%
  \ifdim \wd\@tempboxa >\hsize
    {\small \textbf{#1.}\ \ #2}\par \else
    \global \@minipagefalse
    \hb@xt@\hsize{\hfil\box\@tempboxa\hfil}%
  \fi \vskip\belowcaptionskip}

\def\numberwithin#1#2{\@ifundefined{c@#1}{\@nocounterr{#1}}{%
  \@ifundefined{c@#2}{\@nocnterr{#2}}{%
  \@addtoreset{#1}{#2}%
  \toks@\@xp\@xp\@xp{\csname the#1\endcsname}%
  \@xp\xdef\csname the#1\endcsname
    {\@xp\@nx\csname the#2\endcsname.\the\toks@}}}}
\def\E^#1{{\buildrel #1 \over\vee}}
\newtheorem{theorem}{Theorem}

{\theoremstyle{definition}

}

\begin{document}

\FirstPageHeading{I.V. Gapyak, V.I. Gerasimenko}

\ShortArticleName{On origin of the Fokker -- Planck kinetic evolution}

\AuthorNameForHeading{I.V. Gapyak, V.I. Gerasimenko}

\ArticleName{On Microscopic Origin of the Fokker -- Planck \\ Kinetic Evolution of Hard Spheres}

\Author{I.V. Gapyak$^\ast$\footnote{E-mail: \emph{gapjak@ukr.net}}
        and V.I. Gerasimenko$^\ast$$^\ast$\footnote{E-mail: \emph{gerasym@imath.kiev.ua}}}

\Address{$^\ast$\hspace*{2mm}Taras Shevchenko National University of Kyiv,\\
    \hspace*{4mm}Department of Mechanics and Mathematics,\\
    \hspace*{4mm}2, Academician Glushkov Av.,\\
    \hspace*{4mm}03187, Kyiv, Ukraine}

\Address{$^\ast$$^\ast$Institute of Mathematics of NAS of Ukraine,\\
    \hspace*{4mm}3, Tereshchenkivs'ka Str.,\\
    \hspace*{4mm}01601, Kyiv-4, Ukraine}

\bigskip

\Abstract{The rigorous approach to the description of the kinetic evolution of a many-particle
system composed of a trace hard sphere and an environment of finitely many hard spheres is developed.
We prove that the evolution of states of a trace hard sphere in an environment can be described within
the framework of the marginal distribution function governed by the generalized Fokker -- Planck kinetic
equation and an infinite sequence of the explicitly defined functionals of this function.}

\bigskip

\vspace{2pc}
\Classification{35Q20; 47J35.}

\makeatletter
\renewcommand{\@evenhead}{
\hspace*{-3pt}\raisebox{-7pt}[\headheight][0pt]{\vbox{\hbox to \textwidth {\thepage \hfil \evenhead}\vskip4pt \hrule}}}
\renewcommand{\@oddhead}{
\hspace*{-3pt}\raisebox{-7pt}[\headheight][0pt]{\vbox{\hbox to \textwidth {\oddhead \hfil \thepage}\vskip4pt\hrule}}}
\renewcommand{\@evenfoot}{}
\renewcommand{\@oddfoot}{}
\makeatother

\newpage
\vphantom{math}

\protect\tableofcontents
\vspace{0.7cm}

\section{Introduction}

The rigorous derivation of kinetic equations for many-particle systems composed
of a trace particle moving in an environment of particles, particularly the Fokker -- Planck
kinetic equation, remains an open problem so far. It should be noted there are wide applications of the
Fokker -- Planck equation to the description of kinetic processes of various nature \cite{Bog,Ch,K,R}.

As is known, the Fokker -- Planck kinetic equation was stated in papers \cite{F},\cite{P} by
the instrumentality of phenomenological treatment. The consistent microscopic derivation of the
Fokker -- Planck equation on the basis of methods of the perturbation theory springs from works of
N.N. Bogolyubov \cite{BK},\cite{B45}. In these works the nature of a stochasticity into deterministic
systems was elucidated for the first time.

In modern research a main approach to the problem of the rigorous derivation of the Fokker -- Planck
kinetic equation lies in the construction of the scaling (diffusion) limit \cite{Sh} of a solution
of evolution equations which describe the evolution of states of a many-particle system composed of
a trace particle and an environment, in particular, a perturbative solution of the corresponding BBGKY
hierarchy \cite{CGP97}. The rigorous results on the justification of the Fokker -- Planck kinetic equation
in scaling limits for particles interacting as hard spheres was obtained in papers \cite{LSCh},\cite{S}.
The review of recent results, including quantum systems, was given in article \cite{Er}.

In this paper we develop a rigorous approach to the description of the kinetic evolution of a many-particle
system composed of a trace hard sphere and an environment of finitely many hard spheres. On the basis of the
stated kinetic cluster expansions of the cumulants of groups of operators, which are the generating evolution
operators of a nonperturbative solution of the BBGKY hierarchy, we prove that all possible states of a trace
hard sphere in an environment at arbitrary moment of time can be described within the framework of the marginal
distribution function of the trace hard sphere governed by the generalized Fokker -- Planck kinetic equation
and the explicitly defined functionals of this function without any approximations.
Thus, we establish that the stated Fokker -- Planck kinetic equation gives an alternative approach
for the description of the evolution of states of a trace particle in an environment. We remark that
the specific Fokker-Planck-type kinetic equations can be derived from the constructed generalized
Fokker-Planck kinetic equation in the appropriate scaling limits or as a result of certain approximations.

We briefly outline the structure of the paper.
In sections 2 we formulate necessary preliminary facts about dynamics of a trace hard sphere in an environment.
In sections 3 the main results related to the origin of the Fokker -- Planck kinetic evolution are stated.
Then in sections 4-6 the main results are proved, in particular in section 5, using the kinetic cluster expansions
of cumulants of operators stated in section 4, we derive the generalized Fokker -- Planck kinetic equation.
Finally, in section 7 we conclude with some observations and perspectives for future research.

\section{The evolution of a trace hard sphere in an environment}

We consider a many-particle system composed of a trace particle and an environment
which is a system of a non-fixed, i.e. arbitrary, but finite number of identical
particles in the space $\mathbb{R}^{3}$. If the environment is in the equilibrium state,
for such a system it is used the term a trace particle in a heat bath (or in a thermostat) \cite{LSCh}.

We assume that particles are elastically interacting hard spheres with a diameter $\sigma>0$.
Let the trace hard sphere with the mass $M$ be characterized by the phase coordinates
$(q,p)\equiv x\in\mathbb{R}^{3}\times\mathbb{R}^{3}$, and identical hard spheres
with the mass $m$ of the environment be characterized by the phase coordinates
$(q_{i},p_{i})\equiv x_{i}\in\mathbb{R}^{3}\times\mathbb{R}^{3},\, i\geq 1$.
For configurations of such a system the set
$\mathbb{W}_{1+n}\equiv\big\{(q,q_1,\ldots,q_n)\in\mathbb{R}^{3(1+n)}\big||q_i-q_j|<\sigma$
for at least one pair  $(i,j):\,i\neq j\in(1,\ldots,n)$ and $|q-q_j|<\sigma$,
if $j\in(1,\ldots,n)\big\}$ is the set of forbidden configurations.

The evolution of all possible states of the trace hard sphere in the environment
is described by the sequence of marginal distribution functions
$F(t)=(1,F_{1+0}(t,x),F_{1+1}(t,x,x_1),\ldots,F_{1+s}(t,$ $x,x_1,\ldots,x_s),\ldots)$,
that satisfy the initial-value problem of the BBGKY hierarchy
\begin{eqnarray}\label{NelBog1}
   &&\frac{\partial}{\partial t}F_{1+0}(t)=\mathcal{L}_{1+0}F_{1+0}(t)+
     \int_{\mathbb{R}^{3}\times\mathbb{R}^{3}}dx_{1}
     \mathcal{L}_{\mathrm{int}}(\mathfrak{t},1)F_{1+1}(t),\\
   &&\frac{\partial}{\partial t}F_{1+s}(t)=\mathcal{L}_{1+s}F_{1+s}(t)+
     \int_{\mathbb{R}^{3}\times\mathbb{R}^{3}}dx_{s+1}
     \mathcal{L}_{\mathrm{int}}(\mathfrak{t},s+1)F_{1+s+1}(t)+ \nonumber\\
   &&\hskip+18mm+\sum_{i=1}^{s}\int_{\mathbb{R}^{3}\times\mathbb{R}^{3}}dx_{s+1}
     \mathcal{L}_{\mathrm{int}}(i,s+1)F_{1+s+1}(t),\nonumber\\ \nonumber\\
\label{eq:NelBog2}
   && F_{1+s}(t)_{\mid t=0}=F_{1+s}^0, \quad s\geq0.
\end{eqnarray}
If $t\geq0$, in hierarchy of evolution equations (\ref{NelBog1}) the operator $\mathcal{L}_{1+s}$
is defined by the Poisson bracket of noninteracting particles with the corresponding boundary
conditions on $\partial\mathbb{W}_{1+s}$ \cite{CGP97}:
\begin{equation}\label{OperL}
\begin{split}
   &\mathcal{L}_{1+s}F_{1+s}(t)\doteq-
      \langle \frac{p}{M},\frac{\partial}{\partial q}\rangle_{\mid_{\partial\mathbb{W}_{1+s}}}
      F_{1+s}(t,x,x_1,\dots,x_s)-\\
   &\hskip+21mm-\sum\limits_{i=1}^{s}
      \langle \frac{p_{i}}{m},\frac{\partial}{\partial q_{i}}\rangle_{\mid_{\partial
      \mathbb{W}_{1+s}}}F_{1+s}(t,x,x_1,\dots,x_s),\quad s\geq0,
\end{split}
\end{equation}
where we denote a scalar product by $\langle\cdot,\cdot\rangle$. The operators $\mathcal{L}_{\mathrm{int}}(i,s+1)$
and $\mathcal{L}_{\mathrm{int}}(\mathfrak{t},s+1)$ are defined by the following expressions:
\begin{equation}\label{aLint}
\begin{split}
   &\hskip-3mm\sum_{i=1}^{s}\int_{\mathbb{R}^{3}\times\mathbb{R}^{3}}dx_{s+1}
     \mathcal{L}_{\mathrm{int}}(i,s+1)F_{1+s+1}(t)\doteq\sigma^{2}
     \sum\limits_{i=1}^s\int_{\mathbb{R}^3\times\mathbb{S}_{+}^{2}}d p_{s+1}d\eta\,
     \langle\eta,\big(\frac{p_i}{m}-\frac{p_{s+1}}{m}\big)\rangle\times\\
   &\hskip-1mm\big(F_{1+s+1}(t,x,x_1,\ldots,q_i,p_i^{\star},\ldots,x_s,q_i-\sigma\eta,p_{s+1}^{\star})-
     F_{1+s+1}(t,x,x_1,\ldots,x_s,q_i+\sigma\eta,p_{s+1})\big),\\ \\
   &\hskip-3mm\int_{\mathbb{R}^{3}\times\mathbb{R}^{3}}dx_{s+1}
     \mathcal{L}_{\mathrm{int}}(\mathfrak{t},s+1)F_{1+s+1}(t)\doteq\sigma^{2}
     \int_{\mathbb{R}^3\times\mathbb{S}_{0,+}^{2}}d p_{s+1}d\eta\,
     \langle\eta,\big(\frac{p}{M}-\frac{p_{s+1}}{m}\big)\rangle\times\\
   &\hskip-1mm\big(F_{1+s+1}(t,q,p^*,x_1,\ldots,x_s,q-\sigma\eta,p_{s+1}^{*})-
     F_{1+s+1}(t,x,x_1,\ldots,x_s,q+\sigma\eta,p_{s+1})\big),
\end{split}
\end{equation}
where ${\Bbb S}_{+}^{2}\doteq\{\eta\in\mathbb{R}^{3}\big|\,|\eta|=1,\langle\eta,(p_i-p_{s+1})\rangle>0\}$
and $\mathbb{S}_{0,+}^{2}\doteq\{\eta\in\mathbb{R}^{3}\big|\,|\eta|=1,\langle\eta,(\frac{p}{M}-\frac{p_{s+1}}{m})\rangle>0\}$.
The momenta $p_{i}^{\star}$, $p_{s+1}^{\star}$ and $p^{*}$, $p_{s+1}^{*}$ are defined by equalities:
\begin{equation} \label{eq:momenta}
\begin{split}
   &p_i^{\star}\doteq p_i-\eta\,\left\langle\eta,\left(p_i-p_{s+1}\right)\right\rangle, \\
   &p_{s+1}^{\star}\doteq p_{s+1}+\eta\,\left\langle\eta,\left(p_i-p_{s+1}\right)\right\rangle;\\ \\
   &p^{*}\doteq p-\frac{2Mm}{M+m}\eta\,\langle\eta,\big(\frac{p}{M}-\frac{p_{s+1}}{m}\big)\rangle, \\
   &p_{s+1}^{*}\doteq p_{s+1}+\frac{2Mm}{M+m}\eta\,\langle\eta,\big(\frac{p}{M}-\frac{p_{s+1}}{m}\big)\rangle .
\end{split}
\end{equation}
If $t\leq0$, the BBGKY hierarchy generator is defined by the corresponding operator \cite{CGP97}.

Further we consider initial data (\ref{eq:NelBog2}) of statistically independent a trace hard sphere
and hard spheres of an environment, i.e. at initial instant the marginal distribution functions satisfy
the condition (a chaos property \cite{CGP97})
\begin{equation}\label{eq:Bog2_haos}
\begin{split}
    &F_{1+s}(t)_{\mid t=0}=F_{1+0}^0(x)F^0_{0+s}(x_1,\ldots,x_s)\prod_{i=1}^{s}\mathcal{X}_{2}(q,q_i),\quad s\geq0,
\end{split}
\end{equation}
where $\mathcal{X}_2(q,q_i)$ is the Heaviside step function of allowed configurations $\mathbb{R}^{6}\setminus \mathbb{W}_2$.

To construct a solution of initial-value problem (\ref{NelBog1})-(\ref{eq:Bog2_haos}) we shall adduce
some preliminaries about dynamics of the examined system.
Let $L^1_{1+n}\equiv L^1(\mathbb R^{3(1+n)}\times(\mathbb R^{3(1+n)}\setminus \mathbb{W}_{1+n}))$
be the space of integrable functions $f_{1+n}$ defined on the phase space of $1+n$ particles
that are symmetric with respect to the permutations of the arguments $x_1,\ldots,x_n$, nonsymmetric
with respect to the permutations of the argument $x$ and the arguments $x_1,\ldots,x_n$, and equal to
zero on the set of forbidden configurations $\mathbb{W}_{1+n}$. We denote by $L_{1+n,0}^1\subset L^1_{1+n}$
the subspace of continuously differentiable functions with compact supports.

On a set of measurable functions $f_{1+n}$ defined on the phase space
$\mathbb R^{3(1+n)}\times(\mathbb R^{3(1+n)}\setminus \mathbb{W}_{1+n})$
the following one-parameter mapping: $\mathbb{R}\ni t\mapsto S_{1+n}(-t)f_{1+n}$, is defined by the formula:
\begin{eqnarray} \label{Sspher}
    &&\hskip-5mmS_{1+n}(-t,\mathfrak{t},1,\ldots,n)f_{1+n}(x,x_1,\ldots,x_{n})\doteq\\
    &&\hskip-5mm\doteq\begin{cases}
       f_{1+n}\big(\texttt{X}(-t,x,x_1,\ldots,x_{n}),\texttt{X}_{1}(-t,x,x_1,\ldots,x_{n}),\ldots,\texttt{X}_{n}(-t,x,x_1,\ldots,x_{n})\big),
       \\ \qquad\qquad\qquad\qquad\qquad (x,x_1,\ldots,x_{n})\in(\mathbb R^{3(1+n)}\times(\mathbb R^{3(1+n)}\setminus \mathbb{W}_{1+n}))
       \setminus{\mathcal{M}}_{1+n}^0,\\0,\qquad\qquad\qquad\qquad\,\,\,\,\,\,\,(q,q_{1},\ldots,q_{n})\in \mathbb{W}_{1+n},
    \end{cases}\nonumber
\end{eqnarray}
where $\texttt{X}_{i}(-t)$ is a phase trajectory of $ith$ hard sphere of an environment
and $\texttt{X}(-t)$ is a phase trajectory of a trace hard sphere constructed in \cite{CGP97}.
We note that phase trajectories of a hard sphere system are defined almost everywhere on the
phase space $\mathbb R^{3(1+n)}\times(\mathbb R^{3(1+n)}\setminus \mathbb{W}_{1+n})$ beyond
the set $\mathbb{M}_{1+n}^0$ of the zero Lebesgue measure \cite{CGP97}.

A generator of the isometric group of operators (\ref{Sspher}) is the Liouville operator defined by (\ref{OperL}).

If $F_{1+0}^0\in L^{1}(\mathbb R^{3}\times \mathbb R^{3})$ and
$F_{0+s}^0\in L^{1}(\mathbb R^{3s}\times(\mathbb R^{3s}\setminus \mathbb{W}_{s}))$,
then a nonperturbative solution of initial-value problem (\ref{NelBog1}),(\ref{eq:Bog2_haos})
is a sequence of the distribution functions $F_{1+s}(t,x,x_1,\ldots,x_s),\,s\geq0$,
represented by the following series:
\begin{equation}\label{F(t)}
\begin{split}
   F_{1+s}(t,x,x_1,\ldots,x_{s})&=\sum\limits_{n=0}^{\infty}\frac{1}{n!}\int_{(\mathbb{R}^{3}\times\mathbb{R}^{3})^{n}}
      dx_{s+1}\ldots dx_{s+n}\,\mathfrak{A}_{1+n}(-t,\{\mathfrak{t},Y\},X\setminus Y)F_{1+0}^0(x)\times\\
   &\times F^0_{0+s+n}(x_1,\ldots,x_{s+n})\prod_{i=1}^{s+n}\mathcal{X}_2(q,q_i),
\end{split}
\end{equation}
where the generating evolution operator  $\mathfrak{A}_{1+n}(-t)$ is the $(n+1)th$-order cumulant of groups
of operators (\ref{Sspher}):
\begin{equation}\label{nLkymyl}
\begin{split}
   &\mathfrak{A}_{1+n}(-t,\{\mathfrak{t},Y\},X\setminus Y)=\sum\limits_{\texttt{P}:\,(\{\mathfrak{t},Y\},\,X\setminus Y)
     ={\bigcup\limits}_i X_i}(-1)^{|\texttt{P}|-1}(|\texttt{P}|-1)!\prod_{X_i\subset \texttt{P}}S_{|\theta(X_i)|}(-t,\theta(X_i)),
\end{split}
\end{equation}
and the following notations are used: $\{\mathfrak{t},Y\}$ is a set consisting of one element
$(\mathfrak{t},Y)$, де $Y\equiv(1,\ldots,s)$, i.e. $|\{\mathfrak{t},Y\}|=1$, $\sum_\texttt{P}$
is a sum over all possible partitions $\texttt{P}$ of the set
$(\{\mathfrak{t},Y\},X\setminus Y)\equiv(\{\mathfrak{t},Y\},s+1,\ldots,s+n)$
into $|\texttt{P}|$ nonempty mutually disjoint subsets $X_i\in(\{\mathfrak{t},Y\},X\setminus Y)$,
the mapping $\theta$ is the declusterization mapping defined by the formula: $\theta(\{Y\},X\setminus Y)=X$.

Let $F_{1+0}^0\in L^1(\mathbb{R}^{3}\times\mathbb{R}^{3})$ and initial distribution functions
of an environment belong to the space of integrable functions such that:
$\sup_{n\geq0}\alpha^{-n}\|F^0_{0+n}\|_{L_n^1}$ $<+\infty$, where $\alpha>0$ is a parameter.
Then under the condition that: $\alpha<e^{-1}$, series (\ref{F(t)}) converges in the norm of the space
$L_{1+s}$ for arbitrary $t\in \mathbb{R}^1$. If $F^0_{1+0}\in L^1_{1,0}$ and $F^0_{0+s+n}\in L_{s+n,0}^1$,
the sequence of functions (\ref{F(t)}) is a strong solution of initial-value problem of the BBGKY hierarchy
(\ref{NelBog1}),(\ref{eq:Bog2_haos}) and for arbitrary initial data it is a weak solution \cite{CGP97}.

In consequence of the fact that initial data of a many-particle system composed of a trace particle and an environment
is specified by the initial marginal distribution function of a trace particle, the initial-value problem of the BBGKY
hierarchy (\ref{NelBog1}),(\ref{eq:Bog2_haos}) is not completely well-defined Cauchy problem, because the generic initial
data, is not independent for every unknown marginal distribution function from the hierarchy of evolution equations.
Consequently such initial-value problem can be naturally reformulated as the new Cauchy problem of the kinetic equation
for the marginal distribution function of a trace particle, that corresponds to its initial data, and the sequence
of explicitly defined functionals of a solution of this new Cauchy problem which describe all possible states of
a trace particle and an environment.

\section{The main result: the generalized Fokker -- Planck equation}

In view of the fact that every marginal distribution function of initial data (\ref{eq:Bog2_haos}) is specified 
by the initial marginal distribution function of a trace particle on allowed configurations, the states given 
in terms of the sequence $F(t)=(1,F_{1+0}(t,x),F_{1+1}(t,x,x_1),\ldots,F_{1+s}(t,x,x_1,\ldots,x_s),\ldots)$ of 
marginal distribution functions (\ref{F(t)}) can be described within the framework of the sequence
$F(t\mid F_{1}(t))=(1,F_{1+0}(t,x),F_{1+1}(t,x,x_1\mid F_{1+0}(t)),\ldots,F_{1+s}(t,x,x_1,\ldots,x_s\mid F_{1+0}(t)),\ldots)$
of the marginal functionals of the state $F_{1+s}(t,x,$ $x_1,\ldots,x_s\mid F_{1+0}(t)),s\geq1$, which
are explicitly defined with respect to the solution $F_{1+0}(t,x)$ of the evolution equation for a trace particle.
We refer to such evolution equation for the marginal distribution function of a trace particle as the generalized
Fokker -- Planck kinetic equation.

If $t\geq 0$, the marginal distribution function of a trace particle $F_{1+0}(t,x)$ is the solution of
the generalized Fokker -- Planck kinetic equation
\begin{eqnarray}\label{FPE1}
   &&\hskip-9mm\frac{\partial}{\partial t}F_{1+0}(t,x)=-\langle\frac{p}{M},\frac{\partial}{\partial q}\rangle F_{1+0}(t,x)+\nonumber\\
   &&\hskip-9mm+\sigma^2\sum_{n=0}^{\infty}\frac{1}{n!}\int_{\mathbb{R}^3\times\mathbb{S}^2_{0,+}}d p_1 d\eta
      \int_{(\mathbb{R}^{3}\times\mathbb{R}^{3})^{n}}dx_2\ldots dx_{n+1}\langle\eta,\big(\frac{p}{M}-
      \frac{p_1}{m}\big)\rangle\times\\
   &&\hskip-9mm\times\big(\mathfrak{V}_{1+n}(t,\{\mathfrak{t}^{*},1^{*}_{-}\},2,\ldots,n+1)F_{1+0}(t,q,p^{*})
      -\mathfrak{V}_{1+n}(t,\{\mathfrak{t},1_{+}\},2,\ldots,n+1)F_{1+0}(t,x)\big),\nonumber\\ \nonumber\\
\label{2}
   &&\hskip-9mm F_{1+0}(t,x)|_{t=0}= F_{1+0}^0(x).
\end{eqnarray}
The $(1+n)th$-order generating evolution operator of the collision integral in kinetic equation (\ref{FPE1})
is determined by the expansion (the expansion over scattering cumulants of evolution operators (\ref{nLkymyl})):
\begin{equation}\label{OB2}
  \begin{split}
     &\hskip-5mm\mathfrak{V}_{1+n}(t,\{\mathfrak{t}^{\sharp},1_{\mp}^{\sharp}\},2,\ldots,n+1)F_{1+0}(t,q,p^{\sharp})\doteq\\
     &\doteq n!\,\sum_{k=0}^{n}(-1)^k\,\sum_{m_1=1}^{n}\ldots
       \sum_{m_k=1}^{n-m_1-\ldots-m_{k-1}}\frac{1}{(n-m_1-\ldots-m_k)!}\times\\
     &\times\mathfrak{A}_{1+n-m_1-\ldots-m_k}(t,\{\mathfrak{t}^{\sharp},1^{\sharp}_{\mp}\},2,\ldots,1+n-m_1-\ldots-m_k)\times\\
     &\times F_{0+1+n-m_1-\ldots-m_k}^0(q\mp\sigma\eta,p_1^{\sharp},x_2,\ldots,x_{1+n-m_1-\ldots-m_k})
       \prod_{i_1=2}^{1+n-m_1-\ldots-m_k}\mathcal{X}_2(q,q_{i_1})\times\\
     &\times\mathfrak{A}_1(t,\mathfrak{t}^{\sharp})
       \prod_{j=1}^{k}\big(\frac{1}{m_j!}\,\mathfrak{A}_{1+m_j}(-t,\mathfrak{t}^{\sharp},2+n-m_j-\ldots-m_k,\ldots,
 \end{split}
\end{equation}
\begin{equation*}
  \begin{split}
     &1+n-m_{j+1}-\ldots-m_k)F^0_{0+m_j}(x_{2+n-m_j-\ldots-m_k},\ldots,x_{1+n-m_{j+1}-\ldots-m_k})\times\\
     &\times\prod_{i_2=2+n-m_j-\ldots-m_k}^{1+n-m_{j+1}-\ldots-m_k}\mathcal{X}_2(q,q_{i_2})
       \mathfrak{A}_1(t,\mathfrak{t}^{\sharp})\big)F_{1+0}(t,q,p^{\sharp}),
  \end{split}
\end{equation*}
where the indices $(\mathfrak{t}^{\sharp},1_{\mp}^{\sharp})$ denote that cumulants (\ref{nLkymyl}) of evolution
operators (\ref{Sspher}) act on the phase points $(q,p^{\sharp})$ and $(q\mp\sigma\eta,p_1^{\sharp})$, respectively.

If $t\leq0$, the generalized Fokker -- Planck kinetic equation takes the form:
\begin{eqnarray}\label{FPE2}
   &&\hskip-9mm\frac{\partial}{\partial t}F_{1+0}(t,x)=-\langle\frac{p}{M},\frac{\partial}{\partial q}\rangle F_{1+0}(t,x)+\nonumber\\
   &&\hskip-9mm+\sigma^2\sum_{n=0}^{\infty}\frac{1}{n!}\int_{\mathbb{R}^3\times\mathbb{S}^2_{0,+}}d p_1 d\eta
      \int_{(\mathbb{R}^{3}\times\mathbb{R}^{3})^{n}}dx_2\ldots dx_{n+1}\langle\eta,\big(\frac{p}{M}-
      \frac{p_1}{m}\big)\rangle\times \\
   &&\hskip-9mm\times\big(\mathfrak{V}_{1+n}(t,\{\mathfrak{t},1_{-}\},2,\ldots,n+1)F_{1+0}(t,x)-
      \mathfrak{V}_{1+n}(t,\{\mathfrak{t}^{*},1_{+}^{*}\},2,\ldots,n+1)F_{1+0}(t,q,p^*)\big),\nonumber
\end{eqnarray}
where the generating evolution operators $\mathfrak{V}_{1+n}(t),\,n\geq0,$ are defined by formula (\ref{OB2}).

The marginal functionals of the state $F_{1+s}(t,x,x_1,\ldots,x_s|F_{1+0}(t))$ are represented by the following series:
\begin{equation}\label{f}
\begin{split}
F_{1+s}\big(t,x,x_{1},&\ldots,x_{s}\mid F_{1+0}(t)\big)\doteq\\
    &\doteq\sum_{n=0}^{\infty }\frac{1}{n!}
       \int_{(\mathbb{R}^{3}\times\mathbb{R}^{3})^{n}}dx_{s+1}\ldots dx_{s+n}\,
       \mathfrak{V}_{1+n}(t,\{\mathfrak{t},Y\},X\setminus Y)F_{1+0}(t,x),
\end{split}
\end{equation}
where the generating evolution operators $\mathfrak{V}_{1+n}(t,\{\mathfrak{t},Y\},X\setminus Y),\,n\geq0,$
are defined similar to expansions (\ref{OB2}) and they will be constructed in next section.

We remark that in terms of marginal functionals of the state (\ref{f}) the collision integral
of the generalized Fokker -- Planck kinetic equation (\ref{FPE1}) is represented in the form:
\begin{equation}\label{ci}
\begin{split}
\mathcal{I}_{GFPE}&=\sigma^{2}\int_{\mathbb {R}^3\!\times\mathbb{S}_{0,+}^{2}}
       d p_{1}d\eta\,\langle\eta,\big(\frac{p}{M}-\frac{p_{1}}{m}\big)\rangle \times\\
    &\times\big(F_{1+1}(t,q,p^*,q-\sigma\eta,p_{1}^*\mid F_{1+0}(t))
       -F_{1+1}(t,x,q+\sigma\eta,p_{1}\mid F_{1+0}(t))\big).
\end{split}
\end{equation}
In case of a one-dimensional system the structure of collision integral (\ref{ci}) was considered
in paper \cite{G}.

Thus, the objective of this paper is to prove that initial-value problem of the BBGKY hierarchy
(\ref{NelBog1}),(\ref{eq:Bog2_haos}) is equivalent to initial-value problem of the generalized
Fokker -- Planck equation (\ref{FPE1}),(\ref{2}) and a sequence of marginal functionals of the
state $F_{1+s}\big(t,x,x_{1},\ldots,x_{s}\mid F_{1+0}(t)\big),\,s\geq1$, defined by series (\ref{f}).

Finally, we note that the possibility to describe the evolution of all possible states of
a many-particle system composed of a trace particle and an environment within the framework of
the Cauchy problem of the generalized Fokker -- Planck equation and by a sequence of the marginal
functionals of the state along with the corresponding Cauchy problem of the BBGKY hierarchy is
an inherent property of the description of many-particle systems within the framework of the
formalism of nonequilibrium grand canonical ensemble which is adopted to the description of
infinite-particle systems in suitable functional spaces \cite{CGP97}.

\section{Kinetic cluster expansions of cumulants of evolution operators}

We introduce the transformation of cumulants (\ref{nLkymyl})
which makes possible to represent marginal distribution functions (\ref{F(t)}) in case of $s\geq 1$
in terms of the expansions with respect to the marginal distribution function of a trace hard sphere,
i.e. function (\ref{F(t)}) in case of $s=0$.

We expand cumulants (\ref{nLkymyl}) of operators (\ref{Sspher}) into the following kinetic cluster expansions:
\begin{equation}\label{rrrl2}
\begin{split}
\mathfrak{A}_{1+n}&(-t,\{\mathfrak{t},Y\},X\setminus Y)F_{0+s+n}^0(x_1,\ldots,x_{s+n})
     \prod_{i=1}^{s+n}\mathcal{X}_2(q,q_i)F_{1+0}^0(x)=\\
   &=\sum_{k=0}^{n}\frac{n!}{(n-k)!k!}\mathfrak{V}_{1+n-k}(t,\{\mathfrak{t},Y\},s+1,\ldots,s+n-k)
     \mathfrak{A}_{1+k}(-t,\mathfrak{t},s+n-k+1, \\
   &\ldots,s+n)F_{0+k}^0(x_{s+n-k+1},\ldots,x_{s+n})\prod_{i=s+n-k+1}^{s+n}\mathcal{X}_2(q,q_i)F_{1+0}^0(x),\quad n\geq0,
\end{split}
\end{equation}
where $\mathcal{X}_2(q,q_i)$ is the Heaviside step function of the allowed configurations $\mathbb{R}^{6}\setminus \mathbb{W}_2$
of two hard spheres and $F(0)=(1,F_{1+0}^0(x),\ldots,F_{1+0}^0(x)F^0_{0+s}(x_1,\ldots,x_s),\ldots)$
is the sequence of initial margi\-nal distribution functions (\ref{eq:NelBog2}).
We remark that the structure of cluster expansions (\ref{rrrl2}) is conditioned by
an equivalence of methods of the description of states in terms of a solution of the BBGKY hierarchy (\ref{NelBog1}),
i.e. by the sequence $F(t)=(1,F_{1+0}(t,x),$ $F_{1+1}(t,x,x_1),\ldots,F_{1+s}(t,x,$ $x_1,\ldots,x_s),\ldots)$,
and in terms of the sequence
$F(t\mid F_{1+0}(t))=(1,F_{1+0}(t,x),F_{1+1}(t,x,x_1\mid F_{1+0}(t)),$ $\ldots,F_{1+s}(t,x,$ $x_1,\ldots,x_s\mid F_{1+0}(t)),\ldots)$,
where $F_{1+0}(t)$ is defined by series (\ref{F(t)}) in case of $s=0$, and
$F_{1+s}(t,x,x_1,\ldots,x_s\mid F_{1+0}(t)),\,s\geq1$, are marginal functionals of the state (\ref{f}).

We give a few examples of recurrence relations (\ref{rrrl2})
\begin{equation*}
\begin{split}
   &\mathfrak{A}_{1}(-t,\{\mathfrak{t},Y\})F_{0+s}^0(x_1,\ldots,x_s)\prod_{i=1}^{s}\mathcal{X}_2(q,q_i)F_{1+0}^0(x)=
     \mathfrak{V}_{1}(t,\{\mathfrak{t},Y\})\mathfrak{A}_{1}(-t,\mathfrak{t})F_{1+0}^0(x),\\
   &\mathfrak{A}_{2}(-t,\{\mathfrak{t},Y\},s+1)F_{0+s+1}^0(x_1,\ldots,x_{s+1})\prod_{i=1}^{s+1}\mathcal{X}_2(q,q_i)F_{1+0}^0(x)=\\
   &\hskip+8mm=\mathfrak{V}_{2}(t,\{\mathfrak{t},Y\},s+1)\mathfrak{A}_1(-t,\mathfrak{t})F_{1+0}^0(x)+\\
   &\hskip+8mm+\mathfrak{V}_{1}(t,\{\mathfrak{t},Y\})\mathfrak{A}_{2}(-t,\mathfrak{t},s+1)F^0_{0+1}(x_{s+1})
     \mathcal{X}_2(q,q_{s+1})F_{1+0}^0(x).
\end{split}
\end{equation*}
Solutions of these recurrence relations are given by the following expansions (expansions over scattering operators):
\begin{equation*}
\begin{split}
   &\mathfrak{V}_{1}(t,\{\mathfrak{t},Y\})=\mathfrak{A}_{1}(-t,\{\mathfrak{t},Y\})F_{0+s}^0(x_1,\ldots,x_s)
     \prod_{i=1}^{s}\mathcal{X}_2(q,q_i)\mathfrak{A}_1(t,\mathfrak{t}),\\
   &\mathfrak{V}_{2}(t,\{\mathfrak{t},Y\},s+1)=\mathfrak{A}_{2}(-t,\{\mathfrak{t},Y\},s+1)F^0_{0+s+1}(x_1,\ldots,x_{s+1})
     \prod_{i=1}^{s+1}\mathcal{X}_2(q,q_i)\mathfrak{A}_1(t,\mathfrak{t})-\\
   &\hskip+8mm-\mathfrak{A}_{1}(-t,\{\mathfrak{t},Y\})F_{0+s}^0(x_1,\ldots,x_s)
     \prod_{i=1}^{s}\mathcal{X}_2(q,q_i)\mathfrak{A}_{1}(t,\mathfrak{t})
     \mathfrak{A}_2(-t,\mathfrak{t},s+1)F^0_{0+1}(x_{s+1})\times\\
   &\hskip+8mm \times \mathcal{X}_2(q,q_{s+1})\mathfrak{A}_1(t,\mathfrak{t}).
   \end{split}
\end{equation*}

In the general case solutions of recurrence relations (\ref{rrrl2}), i.e. the $(1+n)th$-order
generating evolution operator $\mathfrak{V}_{1+n}(t)$, is given by the expansion ($s\geq1, n\geq0$):
\begin{equation}\label{OB}
  \begin{split}
   &\mathfrak{V}_{1+n}(t,\{\mathfrak{t},Y\},X\setminus Y)\doteq n!\,\sum_{k=0}^{n}(-1)^k\,\sum_{m_1=1}^{n}\ldots
       \sum_{m_k=1}^{n-m_1-\ldots-m_{k-1}}\frac{1}{(n-m_1-\ldots-m_k)!}\times\\
   &\hskip+5mm\times\mathfrak{A}_{1+n-m_1-\ldots-m_k}(-t,\{\mathfrak{t},Y\},s+1,\ldots,s+n-m_1-\ldots-m_k)\\
   &\hskip+5mm\times F_{0+s+n-m_1-\ldots-m_k}^0(x_1,\ldots,x_{s+n-m_1-\ldots-m_k})
       \prod_{i_1=1}^{s+n-m_1-\ldots-m_k}\mathcal{X}_2(q,q_{i_1})\mathfrak{A}_1(t,\mathfrak{t})\times \\
   &\hskip+5mm\times\prod_{j=1}^{k}\big(\frac{1}{m_j!}\,
       \mathfrak{A}_{1+m_j}(-t,\mathfrak{t},s+1+n-m_j-\ldots-m_k,\ldots,s+n-m_{j+1}-\ldots-m_k)\times\\
   &\hskip+5mm\times F^0_{0+m_j}(x_{s+1+n-m_j-\ldots-m_k},\ldots,x_{s+n-m_{j+1}-\ldots-m_k})
       \prod_{i_2=s+1+n-m_j-\ldots-m_k}^{s+n-m_{j+1}-\ldots-m_k}
       \mathcal{X}_2(q,q_{i_2})\mathfrak{A}_1(t,\mathfrak{t})\big).
  \end{split}
\end{equation}
This statement is proved by induction.

Thus, generating evolution operators (\ref{OB}) of marginal functionals of the state (\ref{f})
and hence the collision integral of the generalized Fokker -- Planck kinetic equation (\ref{FPE1})
are determined by the initial correlations connected with the forbidden configurations of hard spheres
and by the initial state of an environment.

\section{The derivation of the Fokker -- Planck kinetic equation}

Using kinetic cluster expansions (\ref{rrrl2}) of cumulants of operators (\ref{nLkymyl}),
we derive the generalized Fokker -- Planck kinetic equation (\ref{FPE1}) for a trace hard sphere in
an environment which is a system of a non-fixed number of identical hard spheres.

We shall establish that the marginal distribution function defined by series
(\ref{F(t)}),(\ref{nLkymyl}) in case of $s=0$, i.e.
\begin{equation}\label{F1(t)}
\begin{split}
   F_{1+0}(t,x)&=\sum\limits_{n=0}^{\infty}\frac{1}{n!}\int_{(\mathbb{R}^{3}\times\mathbb{R}^{3})^{n}}
      dx_{1}\ldots dx_{n}\,\mathfrak{A}_{1+n}(-t,\mathfrak{t},1,\ldots,n)F_{1+0}^0(x)\times\\
   &\times F^0_{0+n}(x_1,\ldots,x_{n})\prod_{i=1}^{n}\mathcal{X}(q,q_i),
\end{split}
\end{equation}
is governed by evolution equation (\ref{FPE1}) (or (\ref{FPE2})).

In view of the validity in the sense of the norm convergence of the space of integrable functions
of the following equalities for cumulants of groups (\ref{nLkymyl}):
\begin{equation*}
\begin{split}
   &\lim\limits_{t\rightarrow 0}\frac{1}{t}\,\mathfrak{A}_{1}(-t,\mathfrak{t})f_{1+0}(x)
     =\mathcal{L}_{1+0}f_{1+0}(x),\\
   &\lim\limits_{t\rightarrow 0}\frac{1}{t}\int_{\mathbb{R}^3\times\mathbb{R}^3}
     dx_1\,\mathfrak{A}_{2}(-t,\mathfrak{t},1)f_{1+1}(x,x_1)
     =\int_{\mathbb{R}^3\times \mathbb{R}^3}dx_1\mathcal{L}_{\mathrm{int}}(\mathfrak{t},1)f_{1+1}(x,x_1),\\
   &\lim\limits_{t\rightarrow 0}\frac{1}{t}
     \int_{\mathbb{R}^{3n}\times\mathbb{R}^{3n}}
     dx_1\ldots dx_{n}\,\mathfrak{A}_{1+n}(-t,\mathfrak{t},1,\ldots,n)f_{1+n}=0,\quad n\geq2,
\end{split}
\end{equation*}
where $f_{1+0}\in L_0^1(\mathbb{R}^3\times\mathbb{R}^3)$ and $f_{1+n}\in L_0^1(\mathbb{R}^{3(1+n)}\times\mathbb{R}^{3(1+n)})$,
and the operators $\mathcal{L}_{1+0}$ and $\mathcal{L}_{\mathrm{int}}(\mathfrak{t},1)$ are defined by formulas
(\ref{OperL}) and (\ref{aLint}), respectively, then as a result of the differentiation over the time variable of expression (\ref{F1(t)})
in the sense of the pointwise convergence on the space $L^{1}(\mathbb{R}^3\times\mathbb{R}^3)$ we obtain
\begin{equation}\label{de}
\begin{split}
  &\frac{\partial}{\partial t}F_{1+0}(t,x)=
     -\langle \frac{p}{M},\frac {\partial}{\partial q}\rangle F_{1+0}(t,x)+
     \int_{\mathbb{R}^3\times\mathbb{R}^3}dx_1\mathcal{L}_{\mathrm{int}}(\mathfrak{t},1)
     \sum\limits_{n=0}^{\infty}\frac{1}{n!}\int_{(\mathbb{R}^{3}\times\mathbb{R}^{3})^{n}}dx_{2}\\
  &\ldots dx_{n+1}\,\mathfrak{A}_{1+n}(-t,\{\mathfrak{t},1\},2,\ldots,n+1)F_{1+0}^0(x)F_{0+n+1}^0(x_1,\ldots,x_{n+1})
     \prod_{i=1}^{n+1}\mathcal{X}_{2}(q,q_i).
\end{split}
\end{equation}
We represent the second term of the right-hand side of this equality in terms of marginal
distribution function (\ref{F1(t)}) of a trace hard sphere. To this end
we expand cumulants (\ref{nLkymyl}) in series (\ref{F1(t)}) into kinetic cluster expansions (\ref{rrrl2})
for the case $s=1$. Then we transform the series over the summation index $n$ and the sum over the index
$k$ to the two-fold series. As a result the following equality holds:
\begin{equation}\label{scint}
\begin{split}
  &\sum\limits_{n=0}^{\infty}\frac{1}{n!}\int_{(\mathbb{R}^{3}\times\mathbb{R}^{3})^{n}}
     dx_{2}\ldots dx_{n+1}\,\mathfrak{A}_{1+n}(-t,\{\mathfrak{t},1\},2,\ldots,n+1)F_{1+0}^0(x) \times\\
  &\hskip+55mm \times F_{0+n+1}^0(x_1,\ldots,x_{n+1})\prod_{i=1}^{n+1}\mathcal{X}_{2}(q,q_i)=\\
  &=\sum\limits_{n=0}^{\infty}\frac{1}{n!}\sum_{k=0}^{\infty}\frac{1}{k!}
     \int_{(\mathbb{R}^{3}\times\mathbb{R}^{3})^{n+k}}dx_{2}\ldots dx_{n+1+k}
     \mathfrak{V}_{1+n}(t,\{\mathfrak{t},1\},2,\ldots,n+1)\times\\
  &\times\mathfrak{A}_{1+k}(-t,\mathfrak{t},n+2,\ldots,1+n+k)F_{1+0}^0(x)
     F^0_{0+k}(x_{2+n},\ldots,x_{1+n+k})\prod_{i=2+n}^{1+n+k}\mathcal{X}_2(q,q_i).
\end{split}
\end{equation}

According to equalities (\ref{de}) and (\ref{scint}), and taking into account definition (\ref{aLint})
of the operator $\mathcal{L}_{\mathrm{int}}(\mathfrak{t},1)$, from equality (\ref{de}) for $t\geq0$,
we finally derive
\begin{equation*}
  \begin{split}
   &\frac{\partial}{\partial t}F_{1+0}(t,x)=-\langle\frac{p}{M},\frac{\partial}{\partial q}\rangle F_{1+0}(t,x)+\\
   &+\sigma^2\sum_{n=0}^{\infty}\frac{1}{n!}\int_{\mathbb{R}^3\times\mathbb{S}^2_{0,+}}d p_1 d\eta
      \int_{(\mathbb{R}^{3}\times\mathbb{R}^{3})^{n}}dx_2\ldots dx_{n+1}\langle\eta,\big(\frac{p}{M}-
      \frac{p_1}{m}\big)\rangle \times\\
   &\times\big(\mathfrak{V}_{1+n}(t,\{\mathfrak{t}^{*},1^{*}_{-}\},2,\ldots,n+1)
      F_{1+0}(t,q,p^{*})-\mathfrak{V}_{1+n}(t,\{\mathfrak{t},1_{+}\},2,\ldots,n+1)F_{1+0}(t,x)\big),
\end{split}
\end{equation*}
where we used notations accepted in equation (\ref{FPE1}). The collision integral series converges
in the sense of the norm convergence of the space $L^{1}(\mathbb{R}^3\times\mathbb{R}^3)$ under the
condition that: $\alpha<e^{-4}$, where $\sup_{n\geq0}\alpha^{-n}\|F^0_{0+n}\|_{L_n^1}$ $<+\infty$
(the condition on the collision integral coefficients). In next section this fact will be proved
in the general case.

We treat the constructed identity for the marginal distribution function of a trace hard sphere
as the kinetic equation for a trace hard sphere in an environment of identical hard spheres.

Now we consider the structure of the constructed Fokker -- Planck collision integral (\ref{ci}),
namely, we consider the first term of its expansion
\begin{equation*}\label{colint1}
\begin{split}
\mathcal{I}_{GFPE}^{(0)}&=\sigma^2\int_{\mathbb{R}^3\times\mathbb{S}_{0,+}^2}dp_1d\eta \langle\eta,\big(\frac{p}{M}-
     \frac{p_1}{m}\big)\rangle\big(\mathfrak{V}_{1}(t,\{\mathfrak{t}^{*},1^{*}_{-}\})F_{1+0}(t,q,p^*)- \\
  &-\mathfrak{V}_{1}(t,\{\mathfrak{t},1_{+}\})F_{1+0}(t,x)\big).
\end{split}
\end{equation*}
Applying the Duhamel equation to the generating evolution operator $\mathfrak{V}_{1}(t)$
\begin{equation*}\label{Duamel}
\begin{split}
S_2(-t,\mathfrak{t},1)&f_2(x,x_1)=S_1(-t,\mathfrak{t})S_1(-t,1)f_2(x,x_1)+ \\
  &+\int_0^td\tau S_1(-t+\tau,\mathfrak{t})S_1(-t+\tau,1)
     \mathcal{L}_{\mathrm{int}}(\mathfrak{t},1)S_2(-\tau,\mathfrak{t},1)f_2(x,x_1),
\end{split}
\end{equation*}
where the operator $\mathcal{L}_{\mathrm{int}}(\mathfrak{t},1)$ is defined by formula (\ref{aLint}) on
$f_2\in L_{1+1,0}^1$, then the expression $\mathcal{I}_{GFPE}^{(0)}$ is represented in the form:
\begin{equation*}\label{colint12}
\begin{split}
  & \mathcal{I}_{GFPE}^{(0)}=\sigma^2\int_{\mathbb{R}^3\times\mathbb{S}_{0,+}^2}dp_1d\eta \langle\eta,
     \big(\frac{p}{M}-\frac{p_1}{m}\big)\rangle\Big(S_1(-t,1_{-}^*)F^0_{0+1}(q-\sigma\eta,p_1^*)F_{1+0}(t,q,p^*)-\\
  &\hskip+5mm-S_1(-t,1_{+})F^0_{0+1}(q+\sigma\eta,p_1)F_{1+0}(t,x)+
     \int_0^td\tau \big(S_1(-t+\tau,\mathfrak{t}^*)S_1(-t+\tau,1_{-}^*)\times\\
  &\hskip+5mm\times\mathcal{L}_{\mathrm{int}}(\mathfrak{t}^*,1_{-}^*)S_2(-\tau,\mathfrak{t}^*,1_{-}^*)
     F_{0+1}^0(q-\sigma\eta,p_1^*)S_1(t,\mathfrak{t}^*)F_{1+0}(t,q,p^*)-\\
  &\hskip+5mm-S_1(-t+\tau,\mathfrak{t})S_1(-t+\tau,1_{+})
     \mathcal{L}_{\mathrm{int}}(\mathfrak{t},1_{+})S_2(-\tau,\mathfrak{t},1_{+})
     F_{0+1}^0(q+\sigma\eta,p_1)S_1(t,\mathfrak{t})F_{1+0}(t,x)\big)\Big).
\end{split}
\end{equation*}

Thus, the first term of the collision integral $\mathcal{I}_{GFPE}^{(0)}$ of the generalized Fokker -- Planck
equation coincides with the collision integral of the Fokker -- Planck equation established by N.N. Bogolyubov
\cite{Bog} within the framework of the perturbation theory.

We remark that in the space homogeneous case the Markovian approximation of the Fokker -- Planck collision integral
has a more general structure then the canonical collision integral of the Fokker -- Planck equation \cite{R}.

Let $F_{1+0}^0\in L^1(\mathbb{R}^{3}\times\mathbb{R}^{3})$ and the initial distribution functions of an environment
such that $\sup_{n\geq0}\alpha^{-n}\|F^0_{0+n}\|_{L_n^1}$ $<+\infty$,
where $\alpha>0$ is a parameter (it is interpreted as density).
Then for a solution of the Cauchy problem of the generalized Fokker -- Planck equation (\ref{FPE1}),(\ref{2})
in the space of integrable functions $L^{1}(\mathbb{R}^3\times\mathbb{R}^3)$ the following statement is true.
\begin{theorem}
If $\alpha<e^{-4}$, for $t\in\mathbb{R}$ a solution of the Cauchy problem of the generalized Fokker -- Planck
equation (\ref{FPE1}),(\ref{2}) \big((\ref{FPE2}),(\ref{2})\big) is determined by the series:
\begin{equation}\label{ske}
\begin{split}
  &F_{1+0}(t,x)=\\
  &=\sum\limits_{n=0}^{\infty}\frac{1}{n!}\int_{(\mathbb{R}^3\times\mathbb{R}^3)^n}dx_1\ldots dx_{n}\,
     \mathfrak{A}_{1+n}(-t)F_{1+0}^0(x)F_{0+n}^0(x_1,\ldots,x_{n})\prod_{i=1}^{n}\mathcal{X}_{2}(q,q_i),
\end{split}
\end{equation}
where the generating operators $\mathfrak{A}_{1+n}(-t),\,n\geq0,$ are cumulants of groups (\ref{Sspher})
defined by (\ref{nLkymyl}). For initial data $F_{1+0}^0\in{L}^{1}_{0}(\mathbb{R}^3\times\mathbb{R}^3)$
and $F_{0+n}^0\in{L}^{1}_{0}(\mathbb{R}^{3n}\times\mathbb{R}^{3n})$ it is a strong (classical) solution
and for an arbitrary initial data $F_{1+0}^{0}\in L^{1}(\mathbb{R}^3\times\mathbb{R}^3)$ and
$F_{0+n}^0\in{L}^{1}(\mathbb{R}^{3n}\times\mathbb{R}^{3n})$ it is a weak (generalized) solution.
\end{theorem}

The scheme of the proof of this theorem is similar to the proof of an existence theorem for the generalized
Enskog kinetic equation \cite{GG}.

\section{Marginal functionals of the state}

Using kinetic cluster expansions (\ref{rrrl2}), we represent solution expansions (\ref{F(t)}) of
the BBGKY hierarchy (\ref{NelBog1}) in case of $s\geq1$, in the form of the expansions with respect
to marginal distribution function (\ref{F1(t)}) which is governed by the derived Fokker -- Planck
equation (\ref{FPE1}).

In case of $s\geq1$ in every term of series (\ref{F(t)}) we expand cumulants of groups (\ref{nLkymyl})
into kinetic cluster expansions (\ref{rrrl2}). As a result of the transformation of the series over
the summation index $n$ and the sum over the index $k$ to the two-fold series we obtain the following
equality:
\begin{equation*}
\begin{split}
   &F_{1+s}(t,x,x_1,\ldots,x_{s})=
      \sum\limits_{n=0}^{\infty}\frac{1}{n!}\int_{(\mathbb{R}^{3}\times\mathbb{R}^{3})^{n}}
      dx_{s+1}\ldots dx_{s+n}\,\mathfrak{A}_{1+n}(-t,\{\mathfrak{t},Y\},X\setminus Y)F_{1+0}^0(x)\times \\
   &\hskip+5mm\times F^0_{0+s+n}(x_1,\ldots,x_{s+n})\prod_{i=1}^{s+n}\mathcal{X}_2(q,q_i)=\\
   &\hskip+5mm=\sum\limits_{n=0}^{\infty}\frac{1}{n!}\int_{(\mathbb{R}^{3}\times\mathbb{R}^{3})^n}dx_{s+1}\ldots
      dx_{s+n}\,\mathfrak{V}_{1+n}(t,\{\mathfrak{t},Y\},X\setminus Y)
      \sum_{k=0}^{\infty}\frac{1}{k!}\int_{(\mathbb{R}^{3}\times \mathbb{R}^{3})^{k}}dx_{s+n+1}\ldots \\
   &\hskip+5mm \ldots dx_{s+n+k} \mathfrak{A}_{1+n}(-t,\mathfrak{t},s+n+1,\ldots,s+n+k)F^0_{1+0}(x)\times \\
   &\hskip+5mm\times F^0_{0+k}(x_{s+n+1},\ldots,x_{s+n+k})\prod_{i=s+n+1}^{s+n+k}\mathcal{X}_2(q,q_{i}).
\end{split}
\end{equation*}
Taking into account the definition of marginal distribution function (\ref{F1(t)}) of a trace hard sphere
in the obtained expression, finally we establish the equality:
\begin{equation*}
\begin{split}
   &F_{1+s}(t,x,x_1,\ldots,x_{s})=\\
   &=\sum\limits_{n=0}^{\infty}\frac{1}{n!}\int_{(\mathbb{R}^{3}\times\mathbb{R}^{3})^n}dx_{s+1}\ldots
      dx_{s+n}\,\mathfrak{V}_{1+n}(t,\{\mathfrak{t},Y\},X\setminus Y)F_{1+0}(t,x)=\\
   &=F_{1+s}(t,x,x_1,\ldots,x_s|F_{1+0}(t)),\quad s\geq1,
\end{split}
\end{equation*}
where the generating evolution operators $\mathfrak{V}_{1+n}(t),\,n\geq0,$ are defined by formula (\ref{OB})
as solutions of recurrence relations (\ref{rrrl2}).

We establish the existence of marginal functionals of the state (\ref{f}) for
$F_{1+0}^0\in L^1(\mathbb{R}^{3}\times\mathbb{R}^{3})$ and initial data of an environment
such that: $c\equiv\sup_{n\geq0}\alpha^{-n}\|F^0_{0+n}\|_{L_n^1}$ $<+\infty$, where $\alpha>0$
is a parameter which is interpreted as density of an environment.

Owing to the fact that for cumulants of groups (\ref{nLkymyl}) the estimate holds
\begin{equation*}
\begin{split}
  &\int_{(\mathbb{R}^{3}\times\mathbb{R}^{3})^{1+s+n}}dxdx_{1}\ldots dx_{s+n}
      \big|(\mathfrak{A}_{1+n}(-t,\{\mathfrak{t},Y\},X\setminus Y)f_{1+s+n})(x,x_{1},\ldots,x_{s+n})\big|\leq\\ \\
  &\leq n!e^{n+2}\big\|f_{1+s+n}\big\|_{L^{1}_{1+s+n}},
\end{split}
\end{equation*}
then for the $(1+n)th$-order generating evolution operator (\ref{OB}) the following inequality is true:
\begin{equation*}
\begin{split}
   &\int_{(\mathbb{R}^{3}\times\mathbb{R}^{3})^{1+s+n}}dxdx_{1}\ldots dx_{s+n}
      \big|\mathfrak{V}_{1+n}(t,\{\mathfrak{t},Y\},X\setminus Y)F_{1+0}(t,x)\big|\leq
\end{split}
\end{equation*}
\begin{equation*}
\begin{split}
   &\leq n!c^2\alpha^s\big\|F_{1+0}(t)\big\|_{L_1^1}\sum_{k=0}^{n}\sum_{m_1=1}^{n}\ldots\sum_{m_k=1}^{n-m_1-\ldots-m_{k-1}}
      e^{n-m_1-\ldots-m_k+2}\times\\
   &\times \alpha^{n-m_1-\ldots-m_k}\prod_{j=1}^k e^{m_j+2}\alpha^{m_j}=\\
   &=n!c^2\alpha^s\big\|F_{1+0}(t)\big\|_{L_1^1}e^{n+2}\alpha^{n}\sum_{k=0}^{n}e^{2k}
      \sum_{m_1=1}^{n}\ldots\sum_{m_k=1}^{n-m_1-\ldots-m_{k-1}}1.
\end{split}
\end{equation*}
As a result of the validity of this inequality and the following estimate:
\begin{equation*}
\begin{split}
  &\sum_{m_1=1}^{n}\ldots\sum_{m_k=1}^{n-m_1-\ldots-m_{k-1}}1=\frac{(n-k+1)\ldots(n-1)n}{k!}\leq \frac{n^k}{k!}\leq e^n,
\end{split}
\end{equation*}
for marginal functionals of the state (\ref{f}) the estimate holds:
\begin{equation*}
\begin{split}
 &\big\|F_{1+s}\big(t\mid F_{1+0}(t)\big)\big\|_{L^1_{1+s}}\leq
    \big\|F_{1+0}(t)\big\|_{L_1^1}c^2e^2\alpha^{s}\sum_{n=0}^{\infty}e^{2n}\alpha^{n}\sum_{k=0}^{n}e^{2k}=\\
 &=\big\|F_{1+0}(t)\big\|_{L_1^1}c^2e^2\alpha^{s}
    \sum_{n=0}^{\infty}e^{2n}\alpha^{n}\frac{1-e^{2(n+1)}}{1-e^{2}}\leq
    \big\|F_{1+0}(t)\big\|_{L_1^1}c^2e^3\alpha^{s}\sum_{n=0}^{\infty}(e^4\alpha)^n.
\end{split}
\end{equation*}
Hence, functionals (\ref{f}) exist and are represented by converged series provided that:
$\alpha<e^{-4}$.

Thus, in fact we have proved above that marginal distribution functions (\ref{F(t)}) in case of $s\geq1$
and the marginal functionals of the state (\ref{f}) are equivalent if and only if the generating evolution
operators $\mathfrak{V}_{1+n}(t,\{Y\},X\setminus Y),\,n\geq0$, satisfy recurrence relations (\ref{rrrl2}).

We note that the average values of observables are determined by marginal functionals of the state (\ref{f}).
For example, the average value of the $(1+s)$-ary marginal observable
$B^{(1+s)}=(0,\ldots,0,b_{1+s}(x,x_1,\ldots,x_{1+s}),$ $0,\ldots)$ is defined by the formula
\begin{equation*}
\begin{split}
  &\langle B^{(1+s)}\rangle(t)=\\
  &=\frac{1}{(1+s)!}\int_{(\mathbb{R}^{3}\times\mathbb{R}^{3})^{1+s}}
     dxdx_1\ldots dx_s b_{1+s}(x,x_1,\ldots,x_s)F_{1+s}(t,x,x_1,\ldots,x_s\mid F_{1+0}(t)),
\end{split}
\end{equation*}
where the function $F_{1+0}(t,x)$ is a solution of the Cauchy problem of the generalized Fokker -- Planck
equation (\ref{FPE1}),(\ref{2}) \big((\ref{FPE2}),(\ref{2})\big).

We emphasize that in fact constructed functionals of a solution of the generalized Fokker -- Planck
kinetic equation (\ref{f}) characterize all possible correlations which are created in the process
of the evolution a trace hard sphere in an environment.

Thus, in the last two sections we proved the main result of the work, namely, if initial data is
specified by distribution functions (\ref{eq:Bog2_haos}), then all possible states of a trace hard sphere
in an environment at arbitrary moment of time can be described within the framework of marginal distribution
function of a trace hard sphere governed by the generalized Fokker -- Planck equation (\ref{FPE1}) and
the explicitly defined functionals of this function (\ref{f}) without any approximations.

\section{Conclusion}

For a many-particle system composed of a trace hard sphere and an environment
which is a system of a non-fixed number of identical hard spheres we prove an equivalence
of the description of the evolution of states by the Cauchy problem of the BBGKY hierarchy
(\ref{NelBog1}),(\ref{eq:Bog2_haos}) and by the Cauchy problem of the generalized
Fokker -- Planck kinetic equation (\ref{FPE1}),(\ref{2}) and constructed marginal
functionals of its solution (\ref{f}). Thus, the stated Fokker -- Planck kinetic equation
(\ref{FPE1}) is the basis of an alternative approach to the description of the evolution of
a trace particle in an environment.

We remark that in order to describe the evolution of a trace particle in infinite-particle
environment we must to construct a solution of the generalized Fokker -- Planck equation (\ref{FPE1})
for initial data of an environment that belongs to the more general Banach spaces than the space of
integrable functions. In that case every term of solution expansion (\ref{ske}) as well as marginal
functionals of the state (\ref{f}) contains the divergent integrals. The stated structure of generating
evolution operators of mentioned series makes it possible to regularize the corresponding divergent
expressions \cite{G}.

The developed approach is related to the problem of a rigorous derivation of the non-Markovian kinetic equation
from underlaying many-particle dynamics which makes possible to describe the memory effects of the diffusion
processes. The specific Fokker -- Planck-type kinetic equations can be derived from the constructed generalized
Fokker -- Planck kinetic equation in the appropriate scaling limits or as a result of certain approximations.


\addcontentsline{toc}{section}{References}
\renewcommand{\refname}{References}

\end{document}